# Optical vortex array in spatially varying lattice

Amit Kapoor*, Manish Kumar**, P. Senthilkumaran and Joby Joseph

Photonics Research Laboratory, Department of Physics, Indian Institute of Technology Delhi, New Delhi, India 110016

## ABSTRACT

We present an experimental method based on a modified multiple beam interference approach to generate an optical vortex array arranged in a spatially varying lattice. This method involves two steps which are: numerical synthesis of a consistent phase mask by using two-dimensional integrated phase gradient calculations and experimental implementation of produced phase mask by utilizing a phase only spatial light modulator in an optical 4f Fourier filtering setup. This method enables an independent variation of the orientation and period of the vortex lattice. As working examples, we provide the experimental demonstration of various spatially variant optical vortex lattices. We further confirm the existence of optical vortices by formation of fork fringes. Such lattices may find applications in size dependent trapping, sorting, manipulation and photonic crystals.

## 1. INTRODUCTION

Optical Vortex (OV) is an optical wave-field which carries a phase singularity or phase-defect where both real and imaginary values of optical field go to zero[1]. OV has a helical phase variation around its core and its field is given by A(**r**) exp($im\Phi$), where **r** is position vector, $m$ is the topological charge of OV and $\Phi$ is the azimuthal angle around the defect axis. The amplitude variation is such that A(**r**) = 0 at **r** = 0. Each photon of an OV carries an orbital angular momentum of $m\hbar$ [2]. Because of such unique properties OV has been used in wide variety of application like optical testing[3], optical micromanipulation or optical tweezers[5], singular optics[6], optical solitons[7] etc. An isolated OV may be produced by making use of spiral phase plates[8], synthetic holograms[9], liquid crystal cells[10], double cylindrical lens phase converters[11], higher order laser mode separation[12], non-linear optical phenomena[13], dielectric wedge[14] and spiral multiple pin-hole arrangement[15]. In recent times, array of OVs has attracted much attention due to their application in novel types of optical-manipulation and free space optical communications[16,17]. Such an OV array or OV lattice can be produced by variety of interference based methods. Methods based on modified Michelson or Mach-Zehnder interferometer[18,19], pin holes arrangements[20] and phase only spatial light modulators[21,22] are very popular. All the methods produce a uniform OV lattice. Recently, it has been shown that it is possible to generate a spatially varying basis with uniform hexagonal lattice[23]. The different basis in this approach corresponds to different arrangement of OV. It would be highly desirable to be able to control the local orientation and period of the lattice associated with OV array as it would open up possibilities for more control over the trapping and manipulation based applications. But, none of the known methods have shown a spatially dependent or local control on the lattice structure of the generated OV array.

In this paper, we propose a method which enables a simultaneous as well as independent local control over the lattice orientation and lattice period of the OV array. The method for obtaining such local control consists of two simple steps as described by Kumar and Joseph[24]. The steps are: numerical calculation of phase mask and experimental implementation of this phase mask in an optical 4f Fourier filtering setup involving a phase only spatial light modulator. This paper extends the proposed method to do the experimental realization of more complex spatially variant optical vortex lattice. We also confirm the existence of spatially varying optical vortex lattice by interference method where the generated OV array field pattern is interfered with a plane wave to generate fork fringes at each of the vortex locations.

We first describe the numerical synthesis of the phase mask for an OV array spread on a spatially varying hexagonal lattice and then explain the experimental implementation of the same. We also present experimental results showing an independent control over local period and local orientation of OV array spread on a hexagonal lattice.

*Amit.Kapoor@physics.iitd.ernet.in      **manishk.iitd@gmail.com

** Current affiliation: Optics & Photonics Laboratory, EECS Dept., University of Michigan, Ann Arbor, MI 48109-2099

## 2. NUMERICAL SYNTHESIS OF PHASE MASKS

The uniform OV lattice is generated by interference of multiple plane waves where the propagation vectors of all the beams are spread symmetrically along the surface of a cone making same tilt angle with the normal to the interference plane[20]. Such an interference wave-field can be mathematically expressed as

$$\mathbf{E}_{Res}(\mathbf{r}) = \sum_{j=1}^{n} \mathbf{E}_j e^{i(\mathbf{k}_j \cdot \mathbf{r})} . \tag{1}$$

Here the wave-vector $\mathbf{k}_j$ could be expressed in terms of the tilt angle, $\theta$, it makes with normal to the interference plane i.e. z-axis by

$$\mathbf{k}_j = k \times [\{\cos(q_j \pi) \times \sin\theta\}\hat{x} + \{\sin(q_j \pi) \times \sin\theta\}\hat{y} + \cos\theta \hat{z}]. \tag{2}$$

where $k = 2\pi/\lambda$ ($\lambda$ being the wavelength of laser in air) and coefficient $q_j = 2(j-1)/n$.

### 2.1 Phase mask for uniform hexagonal OV lattice

For the simplest uniform OV array spread on a hexagonal lattice we have $n = 3$ in above equation i.e. we effectively have superposition of three plane waves which have the same tilt $\theta$ and are symmetrically distributed in $xy$ plane[4]. The resultant lattice wave-field, formed due to interference of such three unit amplitude plane waves, could be represented by

$$\mathbf{E}_{Res}(\mathbf{r}) = \exp(i\varphi_1) + \exp(i\varphi_2) + \exp(i\varphi_3) , \tag{3}$$

where $\varphi_1$, $\varphi_2$ and $\varphi_3$ are the phase functions and are given by a dot product between the beam wave-vector ($\mathbf{k}_j$) and position vector ($\mathbf{r}$). So, $\varphi_1 = \mathbf{k}_1 \cdot \mathbf{r} = k(\sin\theta)x$, $\varphi_2 = \mathbf{k}_2 \cdot \mathbf{r} = k[(\cos(2\pi/3)\sin\theta)x + (\sin(2\pi/3)\sin\theta)y]$ and $\varphi_3 = \mathbf{k}_3 \cdot \mathbf{r} = k[(\cos(4\pi/3)\sin\theta)x + (\sin(4\pi/3)\sin\theta)y] = k[(\cos(2\pi/3)\sin\theta)x - (\sin(2\pi/3)\sin\theta)y]$. Once $\varphi_1$, $\varphi_2$ and $\varphi_3$ are known it is straight forward to calculate $\mathbf{E}_{Res}$ and then obtain the phase mask by just extracting the phase only component of $\mathbf{E}_{Res}$.

### 2.2 Phase mask for a spatially varying hexagonal OV lattice

For spatially varying lattice the lattice orientation i.e. $\Theta(\mathbf{r})$ and lattice spacing/period i.e. $P(\mathbf{r})$ parameters can be introduced. Calculating a spatially variant lattice wave-vector $\mathbf{K}_j$ according to

$$\mathbf{K}_j(\mathbf{r}) = [\cos(\phi_j + \Theta(\mathbf{r}))\hat{x} + \sin(\phi_j + \Theta(\mathbf{r}))\hat{y}] \times 2\pi/(\lambda \times P(\mathbf{r})), \tag{4}$$

where

$$\phi_j = \cos^{-1}(\mathbf{k}_j \cdot \hat{x}).$$

and using integrated gradient[25] calculations, the phase functions $\varphi_1$, $\varphi_2$ and $\varphi_3$ can be calculated using following equations

$$\nabla\varphi_1(\mathbf{r}) = \mathbf{K}_1(\mathbf{r}) = [\cos(\Theta(\mathbf{r}))\hat{x} + \sin(\Theta(\mathbf{r}))\hat{y}] \times 2\pi/(\lambda \times P(\mathbf{r})), \tag{5}$$

$$\nabla\varphi_2(\mathbf{r}) = \mathbf{K}_2(\mathbf{r}) = [\cos(2\pi/3 + \Theta(\mathbf{r}))\hat{x} + \sin(2\pi/3 + \Theta(\mathbf{r}))\hat{y}] \times 2\pi/(\lambda \times P(\mathbf{r})), \tag{6}$$

and

$$\nabla\varphi_3(\mathbf{r}) = \mathbf{K}_3(\mathbf{r}) = [\cos(2\pi/3 + \Theta(\mathbf{r}))\hat{x} - \sin(2\pi/3 + \Theta(\mathbf{r}))\hat{y}] \times 2\pi/(\lambda \times P(\mathbf{r})). \tag{7}$$

From here, the process of phase mask extraction is exactly same as for a uniform lattice i.e. we make use of eq. (3) to get resultant wave-field and then extract its phase only component to obtain the phase only mask for OV array in spatially varying hexagonal lattice case.

## 3. EXPERIMENTAL SETUP

For the experimental demonstration, we use a diode laser (Toptica-BlueMode, Germany) emitting a 50mW beam at 405nm wavelength, a 20x microscope objective and lenses with focal lengths of 135 mm (for collimation), f = 500 mm (for 4f Fourier filter setup), and a phase-only SLM (Holoeye-LETO, Germany), which is a reflective liquid crystal on silicon microdisplay with 1920x1080 pixels and 6.4μm pixel pitch. Fig. 1 shows the schematic of experimental setup. The numerically computed kinoforms or phase masks were electronically displayed on the SLM. The incident collimated laser beam gets modulated as per the displayed phase mask and generates multiple diffracted beams. There is a non-zero reflection at SLM-air interface which gives rise to the unwanted zero order diffraction term in the Fourier transform plane. A Fourier filter, designed by making use of numerically obtained Fourier transform profile as a guide, was used in the Fourier plane to block all but the desired first order diffracted beams. The lens of 4f setup helps all the beams to come together and interfere to form the desired OV array in the imaging volume. The intensity profile of resultant wave-field was captured on a CMOS camera (DMK-2BUC02, Imaging Source, Germany). Apart from this a detector beam was taken out from original collimated beam using a beam splitter (BS) and was combined with the beams making interference pattern with the help of another BS as shown in Fig. 1. A shutter was introduced in detector beam for blocking/unblocking purpose. Since only one lens was used in 4f arrangement the beams after the lens interfering at CMOS camera plane were diverging beams.

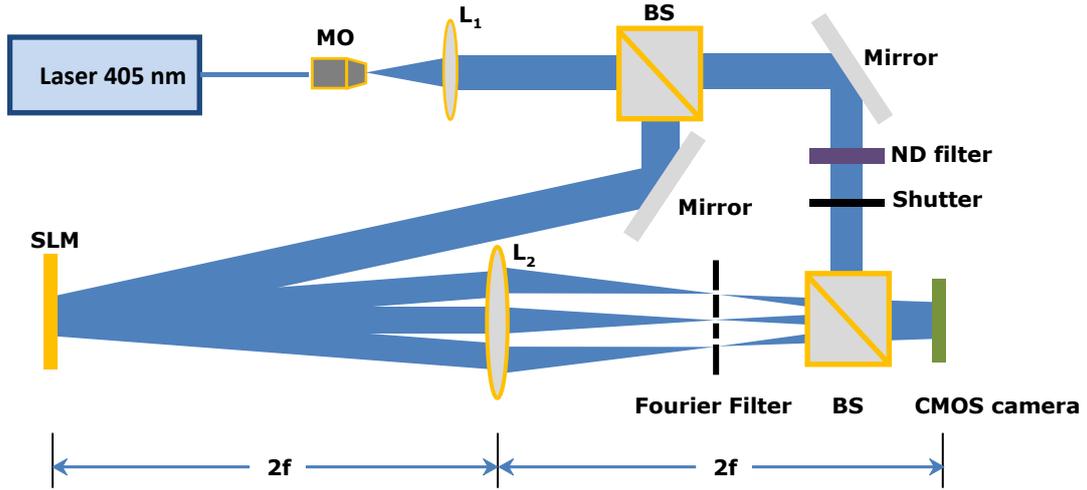

Figure1. Schematic of the experimental setup. MO: Microscope Objective, SLM: Spatial light modulator (phase only), BS: Beam splitter.

## 4. RESULTS AND DISCUSSION

Using the method presented in this paper we generated OV array in hexagonal lattice of varying period and varying orientation. The lattice period variations from left to right, from the center and across the pattern are shown in the simulated patterns of Fig. 2(a), 2(b) and 2(c) respectively. A hexagonal lattice with changing period is shown in Fig. 2d. For obtaining the results in Fig. 2(d), the exact process is followed as described in section 2 of this paper with only change happening as $n = 6$ and resultant wave-field is represented by superposition of waves with modified phase offsets according to the following equation

$$\mathbf{E}_{\text{Res}}(\mathbf{r}) = \sum_{j=1}^{6} \exp(i\varphi_j)\exp(iq_j\pi) . \qquad (8)$$

Figures 2(e) to 2(h) show the experimentally obtained patterns confirming the simulations. Figures 2(i) to 2(l) show simulation results to confirm the existence of vortices in the generated spatially varying vortex arrays and Figure 2(m) to 2(p) are their experimental confirmations.

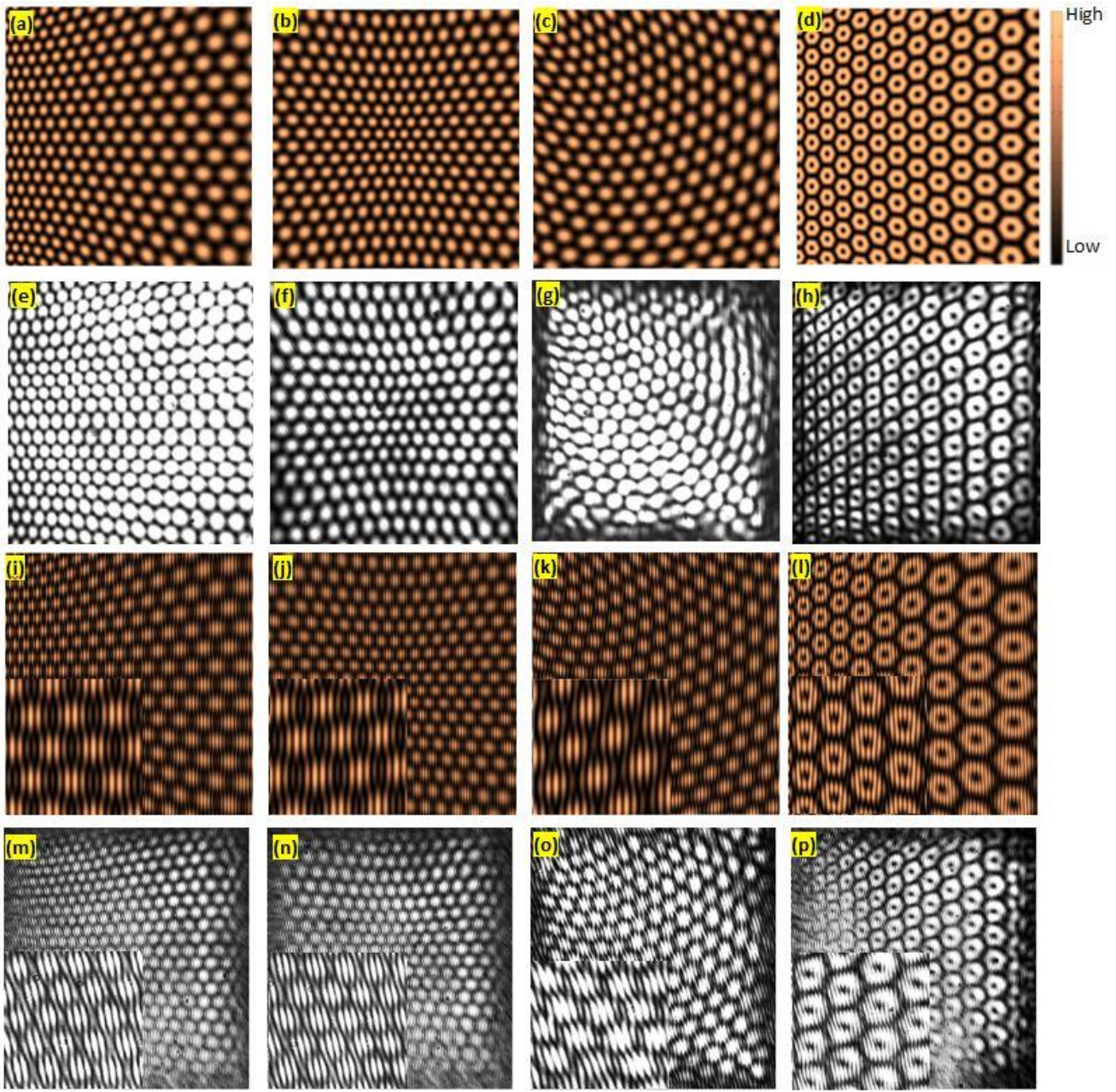

Figure2. Irradiance profiles of various OV arrays in spatially varying lattices. (a) - (d) show simulated irradiance profiles, (e) - (h) show corresponding experimentally obtained result and (i) - (l) show simulations to confirm the existence of optical vortices and (m) - (p) show experimental confirmation of OV. Subsets in (i) to (p) show magnified view of small portion.

The method being a general one could easily be extended to more complex lattice structures as well. The size of the vortices in array can be controlled by controlling the lattice period parameter P(**r**) thereby trapping the objects of different sizes. Trapped objects can further be moved to defined site by controlling the orientation parameter Θ(**r**). It is possible to display multiple phase masks on a SLM at video refresh rate[28] to be able to do a gradual change in the lattice in real time which may be very helpful in particle sorting or size dependent trapping applications.

Figure 3(a) and 3(b) shows the artificial overlap of simulated phase profiles corresponding to case 2(a) and 2(d) respectively with fork fringe pattern of irradiance profile obtained after interference with reference beam. The phase profile has been made transparent to show irradiance profile simulataneously. It can be seen that the places where fork pattern occurs exactly lies with places where optical vortex is present. The respective charges of vortices are also shown in the figure.

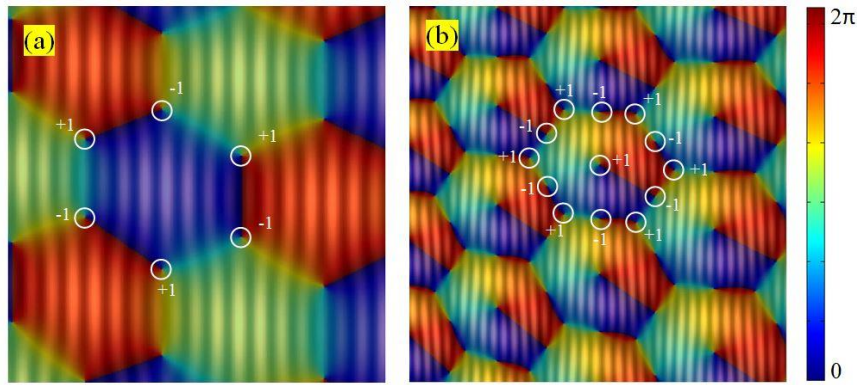

Figure3. Overlapped (artificially) phase and irradiance profiles, (a) spatially varying hexagonal lattice and (b) spatially varying hexagonal vortex lattice.

## 5. CONCLUSIONS

In conclusion, we have provided a simple experimental method for producing spatially varying optical vortex lattice by making use of phase only spatial light modulator assisted 4f Fourier filtering setup. We have also shown experimentally the existence of vortices in lattice array by fork fringes. This work opens up the possibility to have additional degrees of freedom while designing and generating an optical vortex lattice. The method could be very promising for applications in particle trapping and manipulation in real time by controlling the period and orientation parameters and loading the engineered phase masks on SLM at video refresh rates.